\documentclass[12pt]{article}
\usepackage{authblk}
\usepackage[hidelinks]{hyperref}
\usepackage{comment}%
\usepackage[T1]{fontenc}
\usepackage[utf8]{inputenc}
\usepackage{mathpazo}

\usepackage[a4paper, margin=2cm]{geometry}
\usepackage{graphicx}
\usepackage{enumitem}
\usepackage{textcomp}
\usepackage{varwidth}
\usepackage{setspace}
\usepackage[font=normalsize]{caption}
\usepackage{multicol, multirow}
\usepackage[justification=centering]{caption}
\usepackage{color, colortbl, soul}
\usepackage{amsmath}
\usepackage{gensymb}
\usepackage{tabu}
\usepackage{booktabs}
\usepackage{cancel}
\usepackage{cleveref}
\usepackage{listings}
\usepackage{pgfplots}
\usepackage{xcolor}
\usepackage{siunitx}
\usepackage{multicol} %
\usepackage{tabularx}

\usepackage[title]{appendix}

\makeatletter
\renewcommand\appendix{\par
\setcounter{section}{0}%
\setcounter{subsection}{0}%
\setcounter{equation}{0}%
\setcounter{table}{0}%
\setcounter{figure}{0}%
\gdef\theequation{\@Alph\c@section.\arabic{equation}}%
\gdef\thefigure{\@Alph\c@section.\arabic{figure}}%
\gdef\thetable{\@Alph\c@section.\arabic{table}}%
\gdef\thesection{\appendixname\@Alph\c@section}%
\@addtoreset{equation}{section}%
\@addtoreset{table}{section}%
\@addtoreset{figure}{section}%
}
\makeatother

\usepackage[colorinlistoftodos,disable]{todonotes}

\usepackage[modulo]{lineno}

\usepackage{bm}
\DeclareMathAlphabet{\mathsfit}{\encodingdefault}{\sfdefault}{m}{sl}
\SetMathAlphabet{\mathsfit}{bold}{\encodingdefault}{\sfdefault}{bx}{sl}

\newcommand{\up}{\boldsymbol{u}_p}

\newcommand{\upFilt}{\widetilde{\boldsymbol{u}}_p}
\newcommand{\ufFilt}{\widetilde{\boldsymbol{u}}_f}
\newcommand{\ufFiltAtP}{\widetilde{\boldsymbol{u}}_{f@p}}

\newcommand{\jwunit}[1]{\ensuremath{\, \si{#1}}}
\newcommand{\micron}{\ensuremath{\, \si{\micro \metre}}}
\newcommand{\microg}{\ensuremath{\, \si{\micro \gram}}}

\hypersetup{
colorlinks = true,
linkcolor = black, %
citecolor = black, %
urlcolor = blue, %
filecolor = black %
}

\usepackage{natbib}
\setcitestyle{authoryear,open={(},close={)}}

\usepackage{soul}

\usepackage{abstract}

\begin{document}

\doublespacing

\title{Deposition simulations of realistic dosages in patient-specific airways with two- and four-way coupling}
\author[1,2]{Josh Williams}
\author[3]{Jose Manuel Menendez Montes}
\author[4]{Steve Cunningham}
\author[1,5*]{Uwe Wolfram}
\author[1*]{Ali Ozel}

\affil[1]{School of Engineering and Physical Sciences, Heriot-Watt University, Edinburgh, UK}
\affil[2]{STFC Hartree Centre, Daresbury Laboratory, Warrington, UK}
\affil[3]{Information Services, Heriot-Watt University, Edinburgh, UK}
\affil[4]{Centre for Inflammation Research, University of Edinburgh, Edinburgh, UK}
\affil[5]{Institute for Material Science and Engineering, TU Clausthal, Clausthal-Zellerfeld, Germany}
\maketitle
\thanks{* UW and AO share last authorship. \\
Address correspondence to josh.williams@stfc.ac.uk}

\newpage

\begin{abstract}

Inhalers spray over 100 million drug particles into the mouth, where a significant portion of the drug may deposit. 
Understanding how the complex interplay between particle and solid phases influence deposition is crucial for optimising treatments.
Existing modelling studies neglect any effect of particle momentum on the fluid (one-way coupling), which may cause poor prediction of forces acting on particles. In this study, we simulate a realistic number of particles (up to 160 million) in a patient-specific geometry. We study the effect of momentum transfer from particles to the fluid (two-way coupling) and particle-particle interactions (four-way coupling) on deposition. We also explore the effect of tracking groups of particles (`parcels') to lower computational cost. 
Upper airway deposition fraction increased from 0.33 (one-way coupled) to 0.87 with two-way coupling and $10\micron$ particle diameter.
Four-way coupling lowers upper airway deposition by approximately 10\% at $100 \microg$ dosages. 
We use parcel modelling to study deposition of $4 - 20 \micron$ particles, observing significant influence of two-way coupling in each simulation. These results show that future studies should model realistic dosages for accurate prediction of deposition which may inform clinical decision-making.

\end{abstract}
\textbf{Keywords:} Computational particle-fluid dynamics, four-way coupling, image-based model, deposition. \\

\newpage
\listoftodos
\newpage
\section{Introduction}
\label{sec:intro}

Chronic respiratory diseases such as asthma and chronic obstructive pulmonary disease affect over one billion people worldwide \citep{globalAsthma2019}, creating poor patient quality of life and a significant cost burden \citep{nunes2017asthma, chevreul2015costs}. 
These diseases are mainly treated with inhaled drugs delivered with dry-powder inhalers, nebulisers or pressurised metered-dose inhalers. 
Of these, the metered-dose inhaler is the most common. 
Metered-dose inhalers release a large number of aerosol particles over a short period of around $0.1 \jwunit{s}$, with many complex physical interactions such as flash atomisation, particle-particle collisions, cohesion and momentum transfer between drug aerosols and inhaled air.
Patient-specific modelling of metered-dose inhaler drug particle inhalation has been used to study how particles are transported and deposit in the airways \citep{kle07companal, williams2022effect, van2018use}.
As discussed in reviews by \citet{kleinstreuer2010review} and \citet{feng2021tutorial}, the vast majority of existing studies treat aerosols as non-interacting point particles moving through the airways, called one-way coupling (as the drug particles do not influence inhaled air).
This neglects the potential effect of momentum transfer from particles to the fluid phase (two-way coupling), which decreases the kinetic energy of the fluid \citep{elghobashi1994predicting}. Even for low solid volume fraction, but non-zero mass-loading (volume-fraction multiplied by particle-to-fluid density-ratio), the presence of particles can dissipate fluid kinetic energy \citep{boivin1998direct}.
Additionally, particle-particle collisions (four-way coupling) may act to modify the particle trajectories \citep{elghobashi1994predicting}. 
The influence of two-way and four-way coupling on the properties of particle-laden flows have been reviewed extensively \citep{elghobashi1994predicting, balachandar2010review, brandt2022particle}.
Understanding the effect of such interactions is crucial for predicting drug particle deposition statistics in patient-specific airways and optimise patient treatment plans.

Some attempts have been made at investigating the effect of two-way and four-way coupling on drug deposition.
\citet{williams2022effect} performed patient-specific simulations with full coupling between particle and fluid phases, and particle-particle forces such as collisions and cohesion were modelled (four-way coupling). 
However, due to computational limitations it was not possible to model a realistic dosage (i.e. a realistic number of particles) which led to results showing that coupling has a negligible effect.
\citet{xi2022twowaycoupling} performed simulations with a large drug mass injected into a quiescent flow (not an airway geometry), showing the spray plume was altered significantly by two-way coupling effects at moderate and high mass loadings.
However, they used a coarse model which tracks one `parcel' that represents 5000 individual particles.
Simulating this coarser representation could significantly alter the particle transport in gas-solid flows~\citep{benyahia2010estimation}, and sensitivity of the number of particles in a parcel should be investigated.
Similarly, \citet{kolanjiyil2021importance} showed improved prediction of plume velocity and distance travelled with one-way and two-way coupled simulations of nasal spray into a quiescent flow in a similar geometry to \citet{xi2022twowaycoupling}.
However, these analyses did not yet account for four-way coupling effects. 
\citet{kleven2023development} showed the influence of two-way coupling inside an image-based nasal system, showing an improved agreement in deposition measured from nasal casts.
The strong influence of coupling on nasal sprays \citep{kolanjiyil2021importance, kleven2023development} is not surprising due to the large particle size distributions reported (droplet size distribution between $10-100 \micron$ for nasal sprays).
The size distribution is $1-10\micron$ for metered-dose inhalers, which may cause a lower or even negligible influence of coupling due to lower inertia and mass loading of the inhaled particles. 
The significantly smaller particle size used in oral sprays and metered-dose inhalers also imposes a significant computational challenge in modelling studies, since the number of particles for a $100 \microg$ dose of $10 \micron$ particles is approximately 160 million. The computational challenge of simulating more than 100 million particles is a significant barrier to studying the influence of two-way and four-way coupling on drug deposition.

Performing large-scale simulations of realistic numbers of particles is the first step towards understanding the influence of two-way and four-way coupling on drug deposition.
Studies on particle-laden turbulent channel flow in two-way and four-way coupling regime have shown that these effects are highly influential and cannot be neglected.
Specifically, \citet{vance2006properties} showed that collisions cause a significant decrease in the near-wall particle accumulation compared to two-way coupled simulations. 
This occurs as the collisions reduce the particle correlated kinetic energy and increase the random uncorrelated kinetic energy \citep{vance2006properties}. 
As the particles become less correlated with the fluid, they are not driven towards the wall like in one-way and two-way coupled flows.
\citet{longest2010growth} modelled one-way and two-way coupling with condensating droplets in a tubular geometry compared to \textit{in vitro} results, showing improved agreement in predicted aerosol growth (due to condensation) when two-way coupling was included.

Due to the large number of particles in a realistic inhaler payload (more than 100 million), the computational expense of studying a realistic inhaler dosage mass has been bypassed by tracking parcels which represent a population of individual particles \citep{xi2022twowaycoupling, kolanjiyil2021importance, liu2021particle}. 
To ensure the results are valid, it is important to study the influence of the number of particles per parcel (`coarsening factor') against a reference case (where one parcel is one particle). 
This was done by \citet{liu2021particle} when studying cohesive particles dispersion inside a dry-powder inhaler. Using a standard parcel-based coarse-graining approach, they showed the quality of results (fine-particle fraction) degraded when the coarsening factor was larger than 10.
\citet{watanabe2015effects} studied the effect of parcel modelling in a mixing-layer, and showed significant changes in dispersion, momentum transfer and evaporation rate (mass transfer) of the particles when increasing the coarsening factor from 5 to 50. 
Therefore, a sensitivity analysis of the effect of coarsening factor on deposition simulation results will be performed in this study, before embarking on a broader simulation campaign with varying particle size and mass loadings.

In this study, we aimed to first (i) evaluate the influence of two-way and four-way coupling in simulations with a realistic number of drug particles, and then (ii) quantify error incurred from particle-coarsening (parcel-based simulations) on drug deposition in a patient-specific lung airway.

\section{Methods}

\subsection{Mathematical modelling}

Simulations in this study use the Eulerian-Lagrangian approach to model carrier phase (e.g., air) transport and particle dynamics. We solve the volume-filtered incompressible Navier-Stokes equations for the fluid and track point particles by solving Newton's equations of motion.
Simulations were performed in OpenFOAM using a modified MPPICFoam (Multiphase Particle-in-Cell) solver \citep{snider2001}, with modifications described in \autoref{app:code} \citep{codeOCREpaper}.
The transport equations for the fluid phase are given as
\begin{align}
	\frac{\partial}{\partial t}(1-\alpha_p)
    +
    \boldsymbol \nabla \cdot \left[ (1-\alpha_p) \ufFilt \right] 
    &= 0, \label{eq:masscontLES2way} 
    \\
	(1-\alpha_p)
    \left( 
        \frac{\partial \ufFilt }{\partial t} 
        + \ufFilt  \cdot \boldsymbol \nabla \ufFilt  
    \right) 
    &= 
    - \frac{1}{\rho_f}\boldsymbol{\nabla} \widetilde{p}
	+ \frac{1-\alpha_p}{\rho_f}\boldsymbol \nabla \cdot ( \widetilde{\boldsymbol{\tau}} - \boldsymbol{\tau}_{sgs}) 
	+ \frac{1}{\rho_f} \widetilde{\boldsymbol{s}}_{inter}
	+ \boldsymbol{g}.
	\label{eq:momentumEqnLES2way}
\end{align}
where $(\,\widetilde{\cdot}\,)$ represents a volume-filtered field, $\alpha_p$ is the solid phase volume fraction, $\ufFilt$ is the filtered fluid velocity, $\widetilde{p}$ is the filtered fluid pressure and $\widetilde{\boldsymbol{\tau}}$ is the filtered viscous stress tensor, $\boldsymbol{\tau}_{sgs}$ is the subgrid stress tensor \citep{germano91dynamic}, $\rho_f$ is the fluid density, $\boldsymbol{g}$ is gravitational acceleration. The source term for momentum transfer between the fluid and particle phase is $-\boldsymbol{s}_{inter}$. By following \citet{snider2001}, the solid volume fraction for parcels or particles is computed using as
\begin{equation}
    \alpha_p 
    =
    \frac{1}{{V_{cell}}}
    \sum_i^{N_{p,cell}}
    N_{parcel} 
    V_{p(i)} 
    \label{eq:alpp}
\end{equation}
where subscript $(i)$ refers to the \textit{i}th reference particle in the cell, $N_{p,cell}$ is the total number of parcels in a cell, $N_{parcel}$ is the number of particles per parcel, $V_{cell}$ is the volume of a computational cell and $V_p$ is the volume of the reference particle. $N_{parcel}=1$ represents a  ``real'' particle simulation without coarse-graining. 

Momentum transfer between fluid and particle phases is dependent on the sum of the fluid-particle interaction force, $\boldsymbol{f}_{f \rightarrow p}$ (including the effect of $N_{parcel}$). This is computed as $\boldsymbol{f}_{f \rightarrow p(i)}$, averaged over all parcels within a cell volume by, $\boldsymbol{s}_{inter} = - \frac{\sum_{i}^{N_{p,cell}} N_{parcel} \boldsymbol{f}_{f \rightarrow p(i)}}{(1-\alpha_p)V_{cell}}$ \citep{and67fluimech, capecelatro2013euler}.
The two-way coupling regime occurs when $\alpha_p>10^{-6}$ \citep{elghobashi1994predicting} for gas-solid flows, depending also on the density ratio between particle and fluid phases, and local turbulence length-scales \citep{balachandar2010review, brandt2022particle}. 
In the dilute limit ($\alpha_p\rightarrow 0$), there is no influence of the particle motion on the fluid phase or particle-particle interactions (e.g., one-way coupling, $\boldsymbol{s}_{inter}=0$).

The dispersed phase is tracked in the Lagrangian reference frame by following Newton's equations of motion \citep{maxeyriley1983}. We consider only drag (gas-solid flows where the particle density is much larger than the fluid phase density), the gravitational acceleration, the buoyancy force (see Equation Set-III in \cite{zhou2010discrete}) and inter-particle stress \citep{snider2001}. 
The evolution of a parcel velocity, which is equal to the reference particle velocity, $\up$, are computed using the reference particle properties as
\begin{equation}
  m_{p} \frac{d \up}{dt} 
  = 
  \boldsymbol{f}_{f \rightarrow p} 
  + 
  m_p \left( 1 - \frac{\rho_f}{\rho_p} \right) \boldsymbol{g} 
  -
  \frac{m_p}{\rho_{p} \alpha_{p}} \nabla p_p
  \label{eq:up}
\end{equation}
where $m_p$ is the reference particle mass, $\rho_p$ is the reference particle density, and $p_p$ is the solid phase granular pressure.

The drag force acting on a reference particle was computed based on the \citet{wenyudrag} model, given as
\begin{equation}
    \boldsymbol{f}_{f \rightarrow p} = \frac{3}{4} \frac{\mu_f \alpha_f}{d_p^2} Re_p  C_d \alpha_f^{-2.65}  (\ufFiltAtP - \up) 
    \label{eq:forceinteraction}
\end{equation}
where $\mu_f$ is the fluid dynamic viscosity, $\alpha_f$ is the fluid phase volume fraction (equal to $1-\alpha_p$), and $d_p$ is the reference particle diameter. 
The particle Reynolds number is $Re_p = \rho_f \, |\boldsymbol{v}_{r}| \, d_p / \mu_f$, where $\rho_f$ is the fluid density, $|\boldsymbol{v}_{r}|$ is the relative velocity magnitude between the fluid and particle phases. $\ufFiltAtP$ is the filtered fluid velocity at particle position. Using the filtered fluid velocity to compute drag ignores the influence of subgrid fluctuations \citep{williams2022pof, minier2015, innocenti2016}. 
However, models for the subgrid velocity seen by low inertial particles are still in development and therefore we must neglect this until a suitable model is available for non-homogeneous flows. 
$C_d$ is the particle drag coefficient, given as \citep{wenyudrag, schillernaumann}
\begin{equation}
    C_d = \begin{cases}
        \frac{24}{\alpha_f Re_p} \left(1+0.166\left(\alpha_f Re_p\right)^{0.687} \right) &\text{if $\alpha_f Re_p<1000$}, \\
        0.44 &\text{if $\alpha_f Re_p \geq 1000$}.
    \end{cases}
    \label{eq:cd}
\end{equation}
When $\alpha_f=1$, which we explicitly set for one-way coupled simulations, the \citet{wenyudrag} model collapses to that of \citet{schillernaumann}.

When solving particle motion with the discrete element method \citep{cundall79discrete}, forces acting on a particle during collision with another particle are computed with a soft-sphere model based on the overlap distance between two particles. 
This method is highly computationally expensive for aerosol simulations as it requires a timestep $\mathcal{O}(1 - 10 \jwunit{ns})$ for the particle phase \citep{williams2022effect}. 
Also, we have found it is not suitable for a large number of particles (more than one million). 
Therefore, we use the multiphase particle-in-cell (MPPIC) approach to model particle transport with collisions \citep{andrews1996multiphase,snider2001}, where particle properties (volume and velocity) are mapped onto the fluid grid. Particle-phase stresses are then computed using a continuum model, which is then interpolated to the particle position. 
The fluid velocity and fluid volume fraction are also interpolated to particle positions for the drag force calculations (Equations \ref{eq:up}, \ref{eq:forceinteraction} and \ref{eq:cd}). 
Similar to \citet{snider2001}, we used an empricial continuum model for granular solid pressure, $p_p$, to account for the effects of particle-particle collisions with neglecting solid phase shear stress by following \citet{harriscrighton1994}:
\begin{equation}
    p_p = \frac{p^* \alpha_p^ \beta}{\text{max}[\alpha_{cp} - \alpha_p, \epsilon(1-\alpha_p)]} 
    \label{eq:harriscrighton}
\end{equation}
where we use $\beta=2$, $p^*=10 \jwunit{m^2/s^2}$ and $\epsilon=10^{-7}$ \citep{snider2001, verma2020novel}.
Kinetic-theory-based models using the Eulerian random fluctuating kinetic energy (e.g., granular temperature) of particle phase to model the granular solid pressure are available in the literature (see e.g., \citet{lun1984kinetic}). 
In our case the particles or parcels are not uniformly distributed so computing granular temperature to model the granular solid pressure could introduce statistical error in cells with only few particles. For completeness, we include one simulation with the Lun model.

\subsection{Simulation configuration}
All simulations shown used the healthy patient geometry from our previous study \citet{williams2022effect}, also studied by \citet{ban15threedim} and \citet{zhang12sizedepo}. 
We used the same mesh and LES modelling setup that was shown to have excellent agreement with experimental data of fluid phase velocity field in our previous study \citep{williams2022effect, ban15threedim}.
The mesh was a uniform grid with spacing of $\Delta = 500 \micron$. The maximum Courant number was set equal to one which gave a timestep $\Delta t = \mathcal{O}(10 \jwunit{\micro s})$.
In our previous study \citep{williams2022effect} we investigated the effect of number of particles with a constant inlet flowrate that was representative of heavy breathing ($1 \unit{L/s}$). 
In this case, the particles are quickly dispersed in the airways showing a negligible influence of two-way coupling.
In the present study, we used a time-varying inlet boundary condition, as used to compare patient-specific deposition in \citet{williams2022effect}. 
The inhalation profile was obtained from \citet{cola04analtidal} from a healthy patient. The peak flowrate was $0.5 \unit{L/min}$, and followed a near-sinusoidal waveform. 
As the flow is nearly stagnant at the start of inhalation when the patient begins to ramp-up their inhalation, the particles remain entirely in the mouth. 
This causes a higher local solid volume fraction, which is useful for revealing two-way coupling effects.

Since we were focusing on the effect of two-way and four-way coupling, where the main momentum transfer would occur in the mouth and throat, we chose to use a uniform pressure outlet as the distribution of drug in the distal lung was not of interest here.
This also reduced simulation complexity, compared to simulations where a 0D model for outlet pressure based on lung tissue parameters was included \citep{williams2023imaging}.

\begin{figure}
    \centering
    \includegraphics[width=\linewidth]{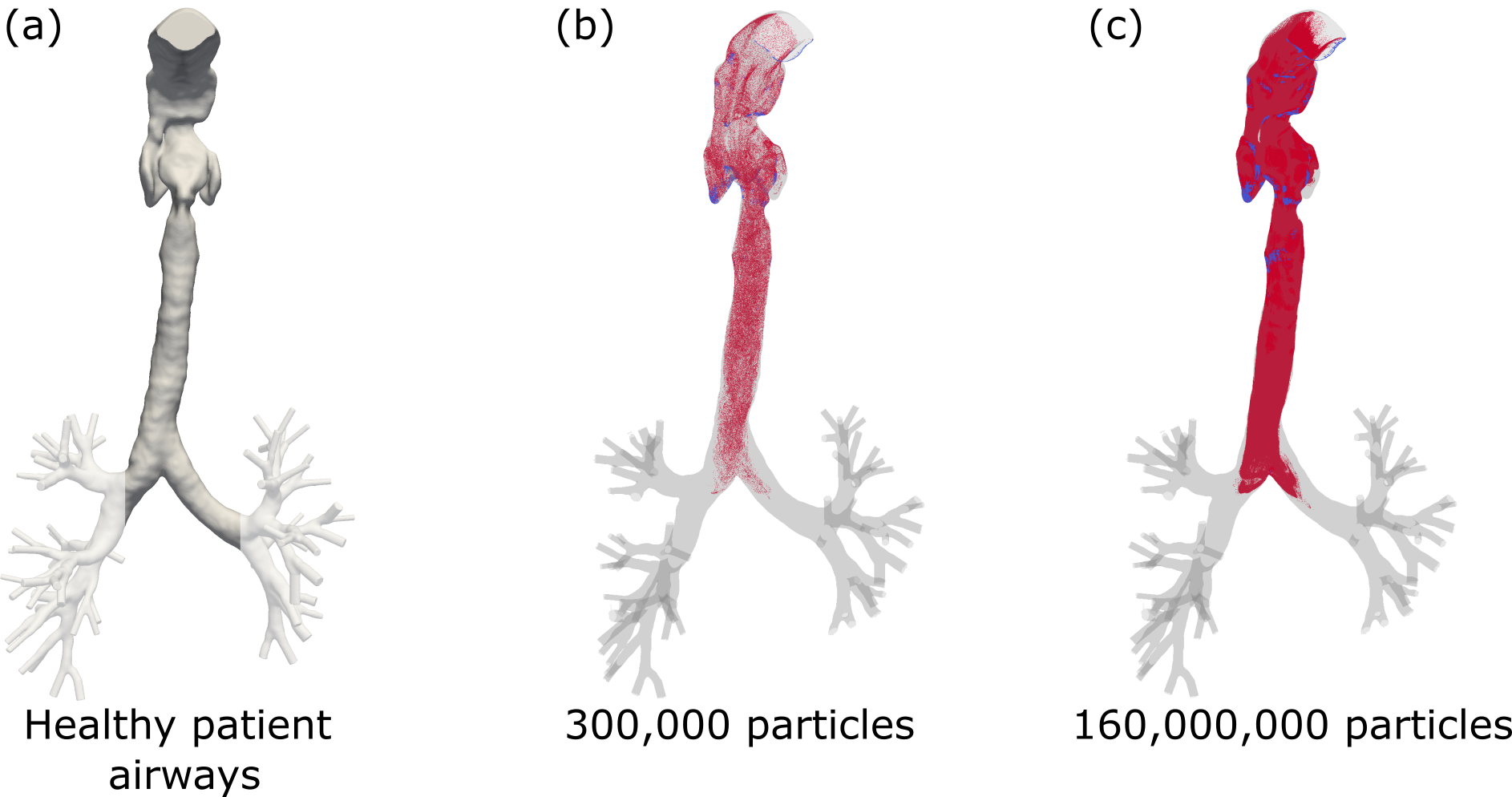}
    \caption{Overview of lung airway geometry considered in this study and deposition visualisations. 
    Panels show (a) segmented healthy patient airways \citep{ban15threedim}, where we have highlighted the `upper airway' region we refer to when discussing regional deposition as used in \citet{williams2022effect}.
    The reminaing panels show visualisations of particle distribution at $t=0.1 \jwunit{s}$ with data from (b) our previous study \citep{williams2022effect} and (c) the present study. }
    \label{fig:introfig}
\end{figure}

We first performed large-scale simulations with a realistic number of particles at $d_p = 10 \micron$, at dosages of $10$, $50$ and $100 \microg$. This corresponded to $15.9 \times 10^6$, $79.2 \times 10^6$ and $159 \times 10^6$ particles (with $N_{parcel}=1$ in Eq.~\eqref{eq:alpp}). The respective computational cost for $1 \unit{s}$ of physical time was 123 CPU-days (50 GB simulation data), 3.35 CPU-years (220 GB simulation data) and 7 CPU-years (19 days with 144 CPU cores, generating 490 GB simulation data). 
We used this simulation data to understand deposition changes with varying dosage. 
In our simulations, we are mainly interested in deposition in the upper airways, shown in \autoref{fig:introfig}a.
We also aim to understand the coupling regime our simulations lie in by comparing the solid phase volume fraction distribution at various time instants \citep{elghobashi1994predicting}.
We also study the correlation between particles and fluid by computing the PDF of alignment between their velocity vectors, $\widetilde{\boldsymbol{n}}_f \cdot \boldsymbol{n}_p$, where $\widetilde{\boldsymbol{n}}_f = \ufFiltAtP / |\ufFiltAtP|$ and $\boldsymbol{n}_p = \up / | \up |$.
To quantify the influence of collisions with varying mass loading, we computed the collision frequency, $\omega_{coll} = \sqrt{\theta} / \lambda_{mfp}$ where $\lambda_{mfp} = d_p / (6 \sqrt{2} \alpha_p)$ is the particle mean-free path \citep{hrenya1997effects}. The granular temperature is $\theta = (1/3) \langle \delta \up \cdot \delta \up \rangle$ \citep{ogawa1980equations}, where $\delta \up = \up - \upFilt$ is the uncorrelated particle velocity. The filtered particle velocity, $\upFilt$, is the local average (Eulerian) particle velocity in each cell.

To reduce the compute time required to model realistic dosages, we then evaluate the effect of parcel-based modelling on the large-scale simulation data. We study the influence of $N_{parcel}=$ 10, 100 and 1000 particles per parcel on deposition in the upper airways, compared to a baseline setup of one particle per parcel. We also study the effect of this coarsening on solid volume fraction distribution, as well as the alignment $\widetilde{\boldsymbol{n}}_f \cdot \boldsymbol{n}_p$.

In the final part of the results, we used a parcel-based model with $N_{parcel}=100$ to explore the effect of mass-loading and two and four-way couplings on deposition at particles in the range $d_p = 4 - 20 \micron$. Again, we study dosages of $10, 50$ and $100 \microg$ for these simulations. These results are compared to one-way coupled simulations, which used a fixed total number of particles, $N_p = 10^6$, with momentum transfer and collisions explicitly deactivated.

\section{Results}

\subsection{Analysis of particle simulations of realistic dosage} \label{sec:results-apriori}
For the case of $d_p = 10 \micron$, we observe a significant increase in deposition compared to one-way coupling at all dosages simulated (see \autoref{tab:10micronDeposition}).
The upper-airway deposition increased by a factor of 1.9, 2.3 and 2.6 by including two-way coupling at $10, 50$ and $100 \microg$, respectively.
At the lowest dosage ($10 \microg$), the deposition fraction only varies by 0.02 when comparing two-way and four-way coupling.
As the dosage is increased, the effect of four-way coupling becomes more significant as the four-way coupled deposition fraction is 18\% smaller than the two-way coupled simulations (relative to the total inhaled dosage). 
Qualitative comparison with deposition visualisations from our previous study \citep{williams2022effect} also show the significantly denser particle concentration with $100 \microg$ ($N_p=200\times 10^6$, \autoref{fig:introfig}b,c).

\begin{table}[!ht]
    \centering
    \caption{Deposition comparison for case of $d_p = 10 \micron$, at various levels of coupling and mass-loading.}
    \begin{tabular}{l c c}
         \toprule
         Simulation & Upper airway deposition fraction & Total deposition fraction \\
         \midrule
         One-way & 0.339 & 0.5364 \\
         \midrule
         Two-way, $10 \microg$ & 0.665 & 0.7408 \\
         Four-way, $10 \microg$ & 0.645 & 0.7288 \\
         \midrule
         
         Two-way, $50 \microg$ & 0.793 & 0.8704 \\
         Four-way, $50 \microg$ & 0.7375 & 0.8036 \\
         \midrule
         
         Two-way, $100 \microg$ & 0.8747 & 0.9075 \\
         Four-way, $100 \microg$ & 0.6974 & 0.7864 \\
         \bottomrule
    \end{tabular}
    \label{tab:10micronDeposition}
\end{table}

\begin{figure}
    \centering
    \includegraphics{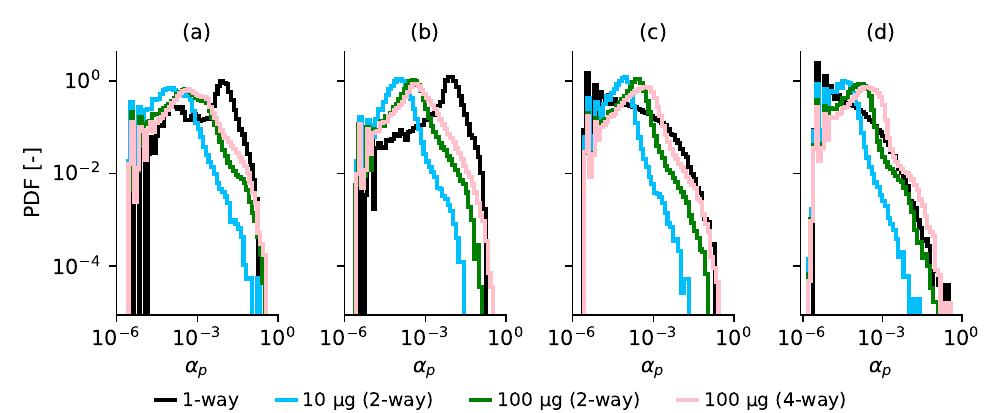}
    \caption{Time evolution of solid volume fraction probability density function (PDF) for $10 \micron$ particles.
    Each plot shows the PDF of solid volume fraction for varying dosages.
    Panels show (a) $t=0.1 \unit{s}$, (b) $t=0.2 \unit{s}$, (c) $t=0.4 \unit{s}$, (d) $t=0.5 \unit{s}$.
    The four-way coupled simulation was performed with the \citet{harriscrighton1994} model for $p_p$.
    The one-way coupled simulation uses the same number of particles as the $10 \microg$ simulation to allow direct comparison of distribution.
    }
    \label{fig:solidVolPDF}
\end{figure}

\begin{figure}
    \centering
    \includegraphics{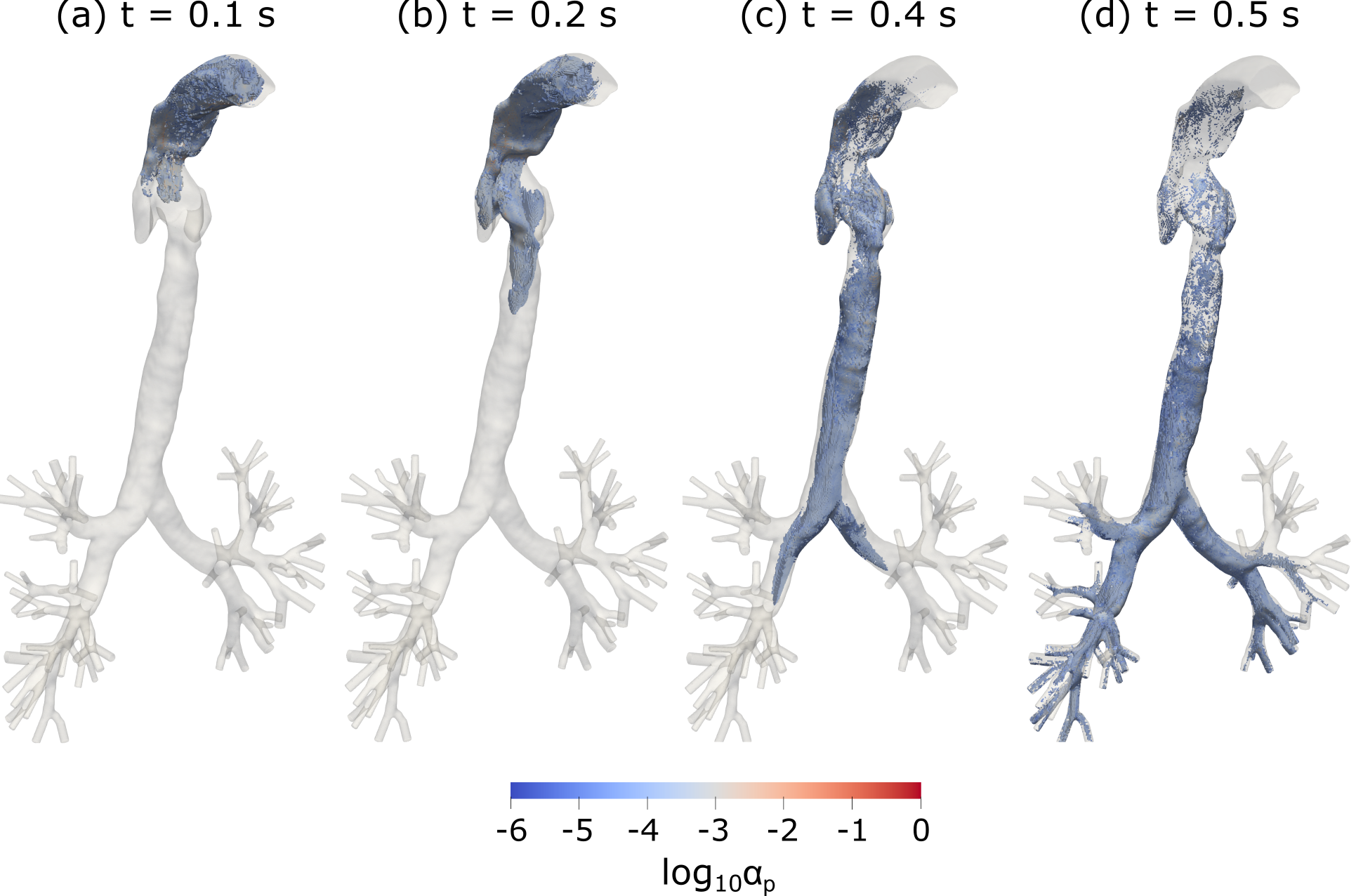}
    \caption{Visualisation of $\log_{10} \alpha_p>10^{-6}$ threshold, corresponding to the $10 \microg$ and $10 \micron$ case with two-way coupling (blue line in \autoref{fig:solidVolPDF}).}
    \label{fig:log10AlphaP-viz}
\end{figure}

We compare the solid volume fraction distribution with varying dosage in \autoref{fig:solidVolPDF}.
Typically one-way coupling is assumed in respiratory drug deposition models \citep{kleinstreuer2010review}, which requires $\alpha_p < 10^{-6}$ \citep{elghobashi1994predicting}. 
From \autoref{fig:solidVolPDF}, we observe that the assumption of one-way coupling does not hold in some regions of the domain that clearly have $\alpha_p \gg 10^{-6}$ (\autoref{fig:solidVolPDF}).
\autoref{fig:log10AlphaP-viz} shows the evolution of $\log_{10} \alpha_p$ with time, highlighting the regions where two-way coupling is influential. Particle concentration in the mouth is dense at the start of the simulation ($t=0.1 - 0.2 \jwunit{s}$). The full airway tree becomes influenced by two-way coupling by $t=0.5 \jwunit{s}$ (\autoref{fig:log10AlphaP-viz}d).
Furthermore, \citet{elghobashi1994predicting} stated that at $\alpha_p > 10^{-3}$, there is an additional contribution of particle-particle interactions (four-way coupling).
As the modelled particle mass approaches a realistic dosage ($100 \microg$), there are some areas of the domain that fall into the four-way coupling regime (\autoref{fig:solidVolPDF}a,b), which helps to explain the influence of four-way coupling in \autoref{tab:10micronDeposition}.
Additionally, the one-way coupled simulation (black line) shows regions of significantly larger solid volume fraction than the two-way coupled case with the same number of particles ($10 \microg$, blue line). 
This suggests that neglecting momentum transfer from particles to the fluid phase in realistic dosage simulations will over-estimate clustering of the particles (discussed later in \autoref{sec:results-deposition}).

\begin{figure}
    \centering
    \includegraphics{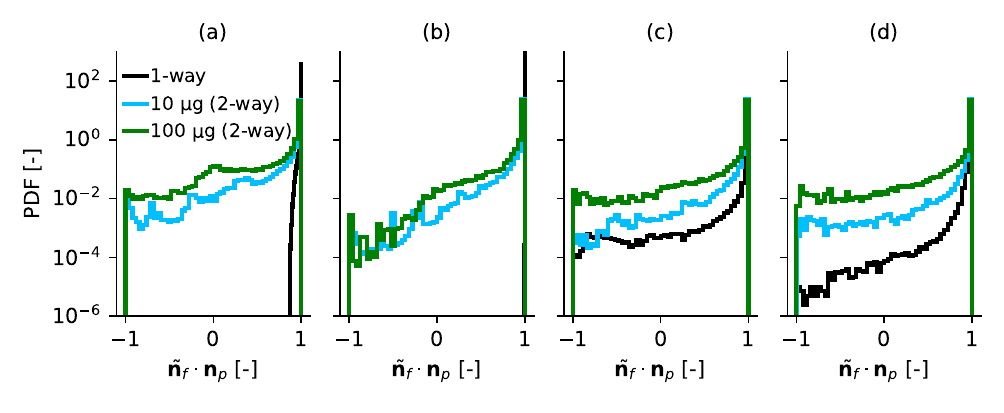}
    \caption{Probability density function (PDF) of alignment between fluid and particle velocity vectors, with various levels of mass-loading and $d_p = 10 \micron$. 
    A value of $\widetilde{\mathbf{n}}_{f} \cdot \mathbf{n}_{p} = 1$ indicates the particle and fluid velocity vectors are fully-aligned.
    Simulations with dosages of $10\microg$ and $100 \microg$ were performed with two-way coupling. 
    Panels show (a) $t=0.1 \jwunit{s}$, (b) $t=0.2 \jwunit{s}$, (c) $t=0.4 \jwunit{s}$, (d) $t=0.5 \jwunit{s}$.
    In panels (a) and (b), nearly all particles in the 1-way coupled PDF are fully-aligned ($\widetilde{\mathbf{n}}_{f} \cdot \mathbf{n}_{p} \rightarrow 1$).
    }
    \label{fig:velocitynormalPDF-10micron}
\end{figure}

To understand how the effect of mass loading modifies particle trajectories, we study the orientation of the particle and fluid velocity vectors, $\widetilde{\mathbf{n}}_{f} \cdot \mathbf{n}_{p}$, at various times throughout the simulation. When this quantity is 1, the fluid and particle velocity vectors are fully aligned. When $\widetilde{\mathbf{n}}_{f} \cdot \mathbf{n}_{p} = 0$, $\up$ is orthogonal to $\ufFilt$. When $\widetilde{\mathbf{n}}_{f} \cdot \mathbf{n}_{p}=-1$, $\up$ is travelling in the opposite direction to $\ufFilt$. At $t=0.1 - 0.2\jwunit{s}$, when all of the particles are released into the domain, the particle velocity is fully aligned with the local fluid velocity in the case of one-way coupling for $d_p = 10 \micron$ (\autoref{fig:velocitynormalPDF-10micron}a,b). 
We then observed that the number of non-aligned particles increases at $t=0.4 \jwunit{s}$ for one-way coupling. This loss of correlation in the one-way coupled case is due to particle inertia, which creates a so-called `crossing-trajectories effect' \citep{yudine1959physical}, where a heavy particle passes through fluid eddies without being perturbed by their motion, causing the particle to quickly become decorrelated with the surrounding fluid and behave less like a tracer particle.
Throughout the simulation, the two-way coupled cases have a more uniform distribution of $\widetilde{\mathbf{n}}_{f} \cdot \mathbf{n}_{p}$ (\autoref{fig:velocitynormalPDF-10micron}a,d), indicating particles are more likely to travel in orthogonal and opposing directions to the local fluid flow.
We observe that two-way coupling and increased mass loading acts to increase the crossing-trajectories effect.

Interestingly, the uniform shape of the two-way coupled velocity PDF is present from the beginning of the simulation (\autoref{fig:velocitynormalPDF-10micron}a, $t=0.1 \jwunit{s}$). 
This occurs as the fluid velocity is near-stagnant and the particle velocity is $|\up^0|=9 \jwunit{m/s}$, causing a strong influence of the momentum transfer source term, $\boldsymbol{s}_{inter}$ on the fluid phase.
Additionally, because the fluid is near-stagnant as the patient begins to inhale, the particles remain concentrated in the mouth-throat region (\autoref{fig:log10AlphaP-viz}a) which creates a large value of $\alpha_p$ (\autoref{fig:solidVolPDF}a).
These factors act to create crossing-trajectories effect observed in \autoref{fig:velocitynormalPDF-10micron}a.
Finally, non-uniformity of the particle initial location across the inlet due to the random initial position acts to modify the fluid velocity gradients through non-uniform $\alpha_p$ and $\boldsymbol{s}_{inter}$, which will cause further misalignment between fluid and particle velocity vectors.

\begin{figure}
    \centering
    \includegraphics{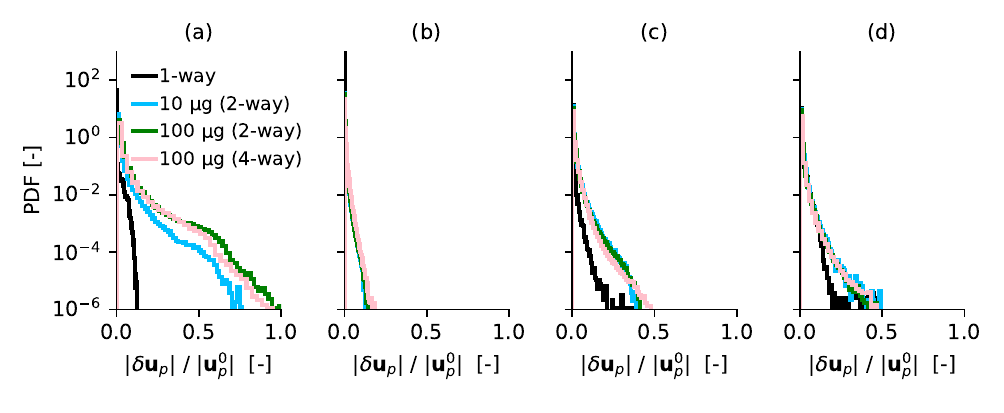}
    \caption{Probability density function (PDF) of uncorrelated velocity magnitude, with various levels of mass-loading and $d_p = 10 \micron$. 
    Simulations with $10\microg$ and $100 \microg$ were performed with two-way coupling. 
    Velocities are normalised by the initial velocity, $|\up^0| = 9 \jwunit{m/s}$.
    Panels show (a) $t=0.1 \jwunit{s}$, (b) $t=0.2 \jwunit{s}$, (c) $t=0.4 \jwunit{s}$, (d) $t=0.5 \jwunit{s}$.
    In panel (b), all PDFs overlap.
    }
    \label{fig:deltaUpPDF-10micron}
\end{figure}

In \autoref{fig:deltaUpPDF-10micron}, we study the effect of varying dosage and coupling on the uncorrelated particle velocity, $\delta \up = \up - \upFilt$.
This was introduced as a means to quantify pseudo-random motion due to particle inertia \citep{fevrier2005partitioning}. At $t=0.1 \jwunit{s}$, we observe high uncorrelated velocity in the two-way and four-way coupled cases compared to the one-way coupled. 
At this early point in the simulation, the one-way coupled particles are mainly following their mean motion which is determined by the initial velocity of the particles.
Later in the simulation ($t = 0.4 - 0.5 \jwunit{s}$, \autoref{fig:deltaUpPDF-10micron}), the one-way coupled $\delta \up$ approaches a similar variance as the two-way and four-way coupled cases. 
At that point, the particle inertia causes some crossing trajectory effects which increases $\delta \up$ (also observed in \autoref{fig:velocitynormalPDF-10micron}).

\begin{figure}%
    \centering
    \includegraphics{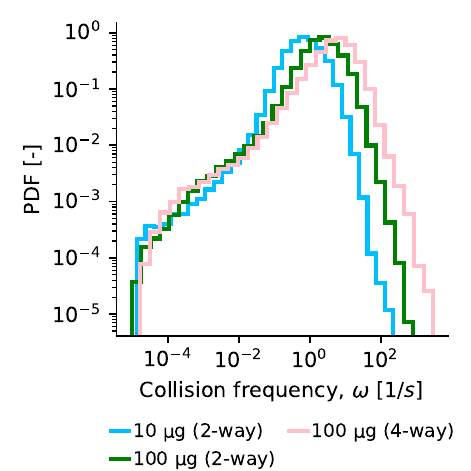}
    \caption{
    Probability density function (PDF) of collision frequency, $\omega = \sqrt{\theta}/\lambda_{mfp}$, with various levels of mass-loading and $d_p = 10 \micron$.
    Data taken at $t = 0.5 \jwunit{s}$.
    }
    \label{fig:collisionfreq-10um}
\end{figure}

As the dosage increases and four-way coupling is accounted for, the collision frequency of the particles increases as a by-product of larger $\alpha_p$ and $\theta$ (\autoref{fig:collisionfreq-10um}).
The low $\omega$ region of the PDF was similar for all three configurations, however the large $\omega$ tail of the PDF increased showing regions of higher collision frequencies.
This is shown by an increase in variance of $\log_{10} \omega$ from 0.26 to 0.32 when increasing $10 \microg$ to $100 \microg$, and a further increase to 0.4 with four-way coupling.

\subsection{Effect of parcel modelling on deposition} \label{sec:results-parcel}

\begin{figure}[!ht]
    \centering
    \includegraphics{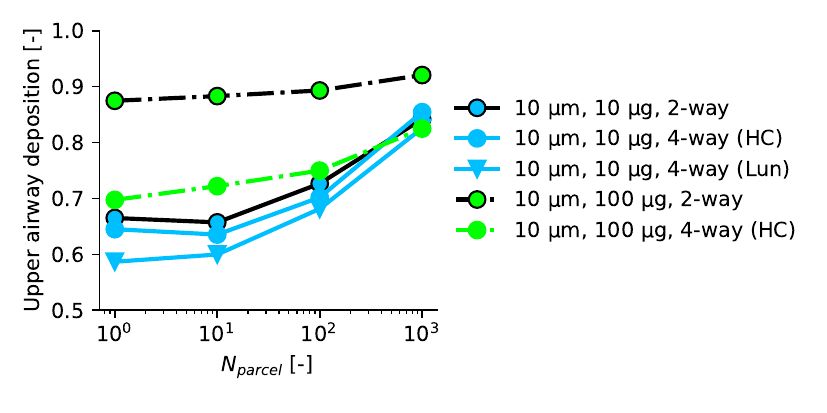}
    \caption{Effect of number of particles per parcel, $N_{parcel}$, on deposition in a representative set of cases. %
    Each face color and line-style represents a different dosage. Two-way coupled lines and markers have black edges, whereas four-way coupled have coloured edges.
    ``HC" and "Lun" refer to simulations with the \citet{harriscrighton1994} and \citet{lun1984kinetic} closures, respectively, for the granular solid pressure, $p_p$. 
    }
    \label{fig:parcel}
\end{figure}

\autoref{fig:parcel} shows the influence of increasing the number of particles per parcel (level of coarsening) on upper airway deposition. 
We anticipated that coarsening the particle phase in this way may over-estimate particle inertia and cause an over-prediction in upper airway deposition, and thus we analyse the sensitivity of this parameter against simulations where each particle is explicitly modelled ($N_{parcel}=1$).
For 10 and 100 particles per parcel, we observed the deposition was always within 10\% of the reference data. 
The maximum error for 10 particles per parcel was 2.5\% for $100 \microg$ with four-way coupling. 
The maximum error for 100 particles per parcel was 9.5\% (relative to total dosage), which was found when using the \citet{lun1984kinetic} model with $10 \microg$ (four-way coupling).

When we increased the number of particles per parcel to 1000, four out of five tests had an error larger than 10\%. 
The maximum error was in the simulation with $10 \microg$ and four-way coupling (Lun model), which gave an error of 23.9\%.
The median absolute error was 0.9\% for $N_{parcel}=10$, 5.7\% for $N_{parcel}=100$ and 17.7\% for $N_{parcel}=1000$.
Clearly, at 1000 particles per parcel the error becomes intolerable and is not suitable for predicting deposition.

\begin{figure}
    \centering
    \includegraphics{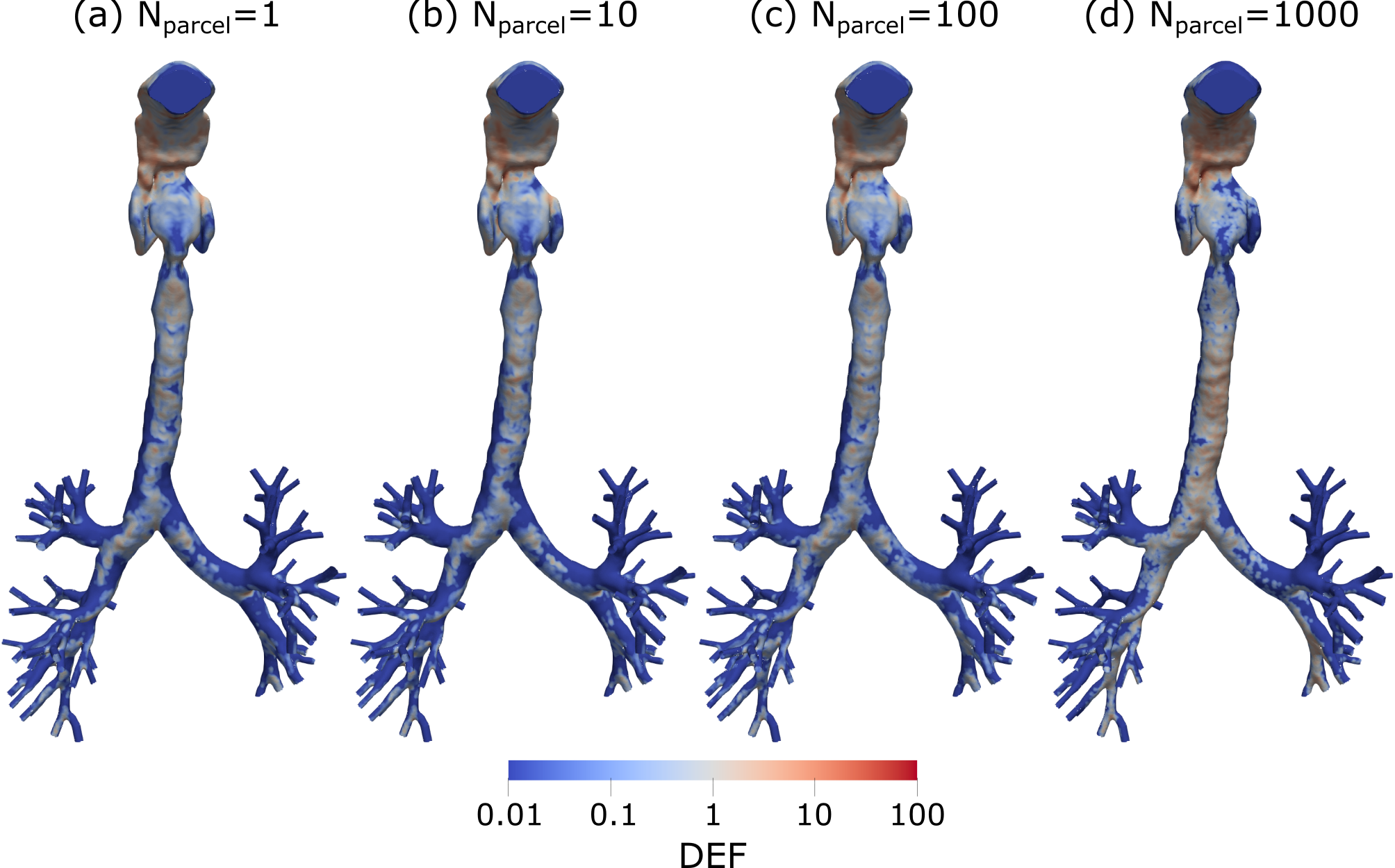}
    \caption{Deposition enhancement factor (DEF), Eq.~\eqref{eq:dosimetry}, of $10 \micron$ particles on the airway wall with various number of particles per parcel, $N_{parcel}$. All simulations use a dosage of $100\microg$ with two-way coupling. Each panels shows an increase in the $N_{parcel}$ by a factor of 10.}
    \label{fig:parcelDEF10um}
\end{figure}

To understand any changes in the local particle concentration due to the use of parcels, we evaluated the deposition enhancement factor (DEF)~\citep{balashazy1999computation,longest2006childhood,williams2022effect} in two cases. The DEF uses the number of particles deposited within a fixed distance ($1 \unit{mm}$, area $A_{conc} = \pi \, (1\unit{mm})^2$~\citep{dong2019numerical}) of the central point of each wall face. Its relation to the global deposition is defined as
\begin{linenomath}
\begin{equation}
	\mathrm{DEF} = \frac{\textrm{Deposition concentration within } A_{conc}/ A_{conc}}{\textrm{Total deposited particles / Total airway surface area }}. \label{eq:dosimetry}
\end{equation}
\end{linenomath}
The first case is the $10\micron$, $100\microg$ two-way coupled simulation shown in \autoref{fig:parcel}, which showed little variation of upper airway deposition. When evaluating the local deposition (\autoref{fig:parcelDEF10um}), we also observe the deposition patterns from $N_{parcel}=1-100$ to be similar. %
At $N_{parcel}=1000$, the deposition in the mouth and trachea is similar, but less smooth due to the lower number of particles giving poorer statistics. In the bronchial airways, it appears deposition is over-estimated, particularly towards the outlets of the airways in the lower lobes.

\begin{figure}
    \centering
    \includegraphics{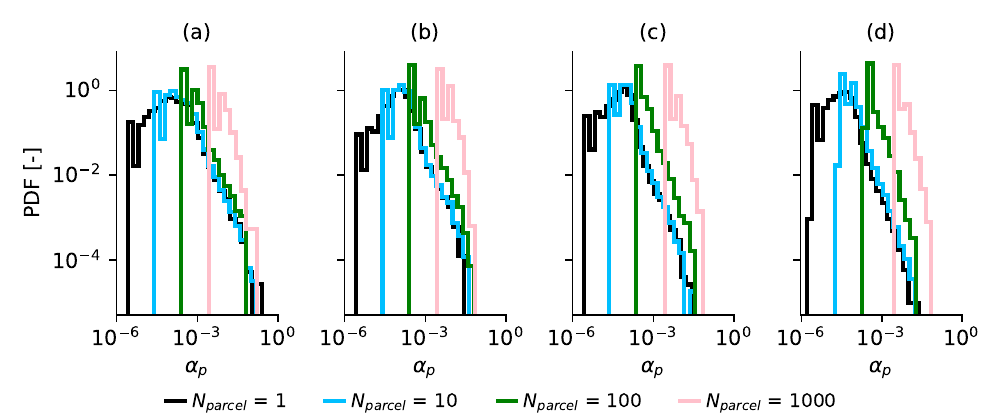}
    \caption{Probability density function (PDF) of solid volume fraction  at various time instants, comparing the effect of $N_{parcel}$. 
    Simulations shown use $d_p = 10 \micron$ and dosage of $10 \microg$ with two-way coupling.
    Panels show (a) $t=0.1 \jwunit{s}$, (b) $t=0.2 \jwunit{s}$, (c) $t=0.4 \jwunit{s}$, (d) $t=0.5 \jwunit{s}$. 
    }
    \label{fig:coarse-solidVolPDF}
\end{figure}

In \autoref{fig:coarse-solidVolPDF}, we study the effect of $N_{parcel}$ on the solid volume fraction distribution to improve our understanding of the deposition increase observed in \autoref{fig:parcel}.
As the particles are non-homogeneously distributed, many cells in the baseline case ($N_{parcel}=1$) contain only one or a few particles. When $N_{parcel}>N_{p,cell}$, this causes the observed shift of the minimum $\alpha_p$ to the right.
The coarsening factor of 10 agrees near exactly with the baseline ($N_{parcel} = 1$) in the right-sided tail of the PDF. This is true throughout the entire simulation (\autoref{fig:coarse-solidVolPDF}a-d). When the coarsening factor is 100, $\alpha_p$ is similar to $N_{parcel} = 10$ in the early parts of the simulation (\autoref{fig:coarse-solidVolPDF}a,b). However, later in the simulation the distribution shows an over-prediction of high $\alpha_p$ regions compared to the baseline (\autoref{fig:coarse-solidVolPDF}). The coarsest model ($N_{parcel}=1000$) shows a large over-prediction of $\alpha_p$ throughout the entire simulation, creating an increase in the effective inertia which explains the over-predicted deposition in \autoref{fig:parcel}.

\begin{figure}
    \centering
    \includegraphics{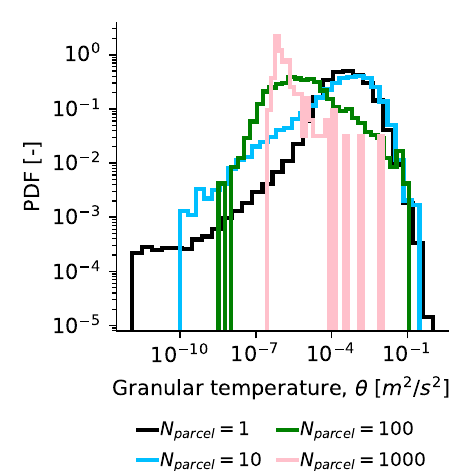}
    \caption{Probability density function (PDF) of granular temperature, comparing the effect of $N_{parcel}$, at $t=0.5\jwunit{s}$.
    Simulations shown use $d_p = 10 \micron$ and dosage of $10 \microg$ with two-way coupling.
    }
    \label{fig:coarse-grantemp}
\end{figure}

\autoref{fig:coarse-grantemp} shows the granular temperature PDF with varying $N_{parcel}$, where the $N_{parcel}=1$ and $N_{parcel}=10$ distributions are in good agreement at high $\theta$.
The $\log_{10} \theta$ variance at $N_{parcel}=100$ was similar to the $N_{parcel}=1$, with respective values of $1.3$ and $0.68$.
However, the $N_{parcel}=1$ and $10$ distribution was more biased towards large $\theta$ values which were not captured by $N_{parcel}=100$.
This is shown by the skewness of the $\log_{10} \theta$ PDF which was -1.02 and -1.27 for $N_{parcel}=1,10$, but 0.6 for $N_{parcel}=100$.
The skewness for $N_{parcel}=1000$ was 3, which is further away from the results with $N_{parcel}=1$.
As the number of particles per parcel increases (total number of parcels decreases), the quality of second-order statistics such as the granular temperature, $\theta$, is degraded.

Results reported in the rest of the paper use $N_{parcel} = 100$, due to the balance of low/moderate deposition error (\autoref{fig:parcel}) and superior computational speedup compared to $N_{parcel}=10$ (we found a speedup of approximately 4 between $N_{parcel}=10$ and $100$). 
This speedup is required to run a broader deposition parameter study with varying dosages and particle sizes (particularly small particles, which have a larger $N_p$ for a given dosage).
The computational speedup gained by varying $N_{parcel}$ will be dependent on the total value of $N_p$, as well as time-taken to compute the fluid fields which will account for a significant portion of compute time when high resolution meshes are used.

\subsection{Effect of mass loading on deposition with parcel approach} \label{sec:results-deposition}

\begin{figure}
    \centering
    \begin{tabular}{c c}
        (a) & (b)  \\
        \includegraphics{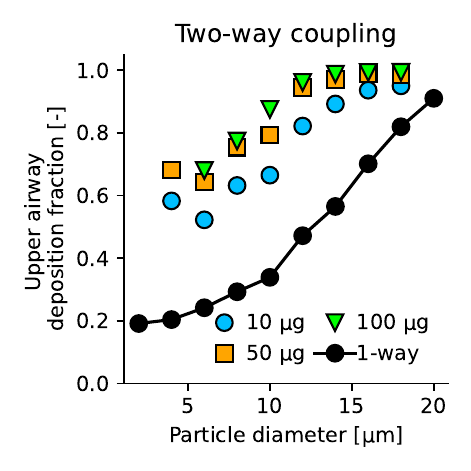} & 
        \includegraphics{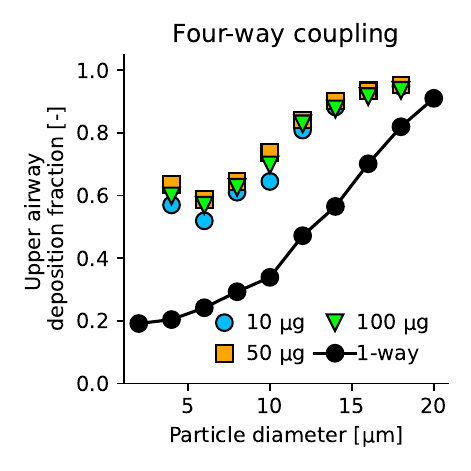} \\
        (c) & (d)  \\
        \includegraphics{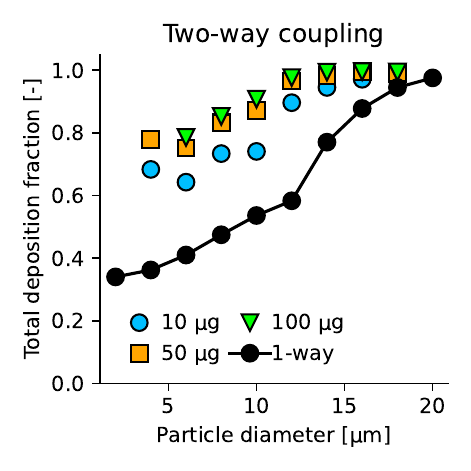} & 
        \includegraphics{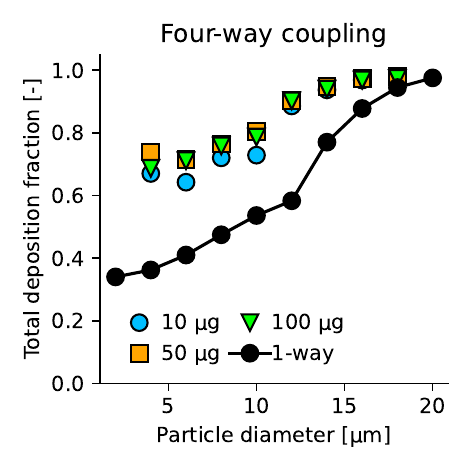}
    \end{tabular}
    \caption{Effect of mass loading levels on deposition in parcel simulations with (a,c) two-way and (b,d) four-way coupling ($N_{parcel}=100$). 
    Top row shows upper airway deposition, bottom row shows total deposition.}
    \label{fig:depositionWithDpMouth}
\end{figure}

We compared the effect of modelled drug mass on upper airway deposition for various particle sizes with $N_{parcel} = 100$ (\autoref{fig:depositionWithDpMouth}). Increasing the drug mass in two-way coupled results increases the deposition compared to one-way coupled simulations (\autoref{fig:depositionWithDpMouth}). As the dosage was increased from $10 \microg$ to $100 \microg$, the deposition continued to increase in two-way coupled simulations (\autoref{fig:depositionWithDpMouth}a). However, in four-way coupled simulations the deposition appears to decrease when the dosage is increased from $50 \microg$ to $100 \microg$. This can be explained by an increased particle granular pressure near the walls due to a higher solid volume fraction, acting to drive particles away from the wall.

\begin{figure}
    \centering
    \begin{tabular}{c c}
        (a) & (b) \\
        \includegraphics{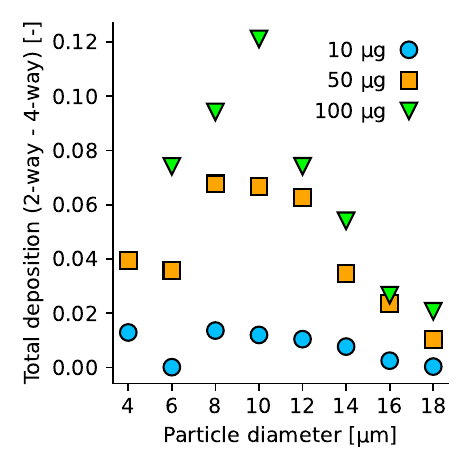} & \includegraphics{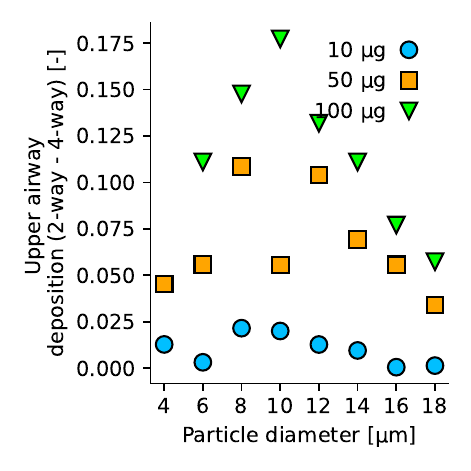} \\
    \end{tabular}
    \caption{Difference between two-way and four-way coupled deposition fractions, based on particle diameter and modelled dose for parcel simulations with ($N_{parcel}=100$). 
    Panels show (a) the total deposition, and (b) deposition in the upper airways.}
    \label{fig:depositionDiff}
\end{figure}

The difference in the deposition of two-way and four-way coupled simulations is shown in \autoref{fig:depositionDiff}. For low doses ($10 \microg$), the difference in two-way and four-way coupling was within 5\% for all particle sizes. 
The influence of four-way coupling is low for larger particles, since the deposition for these particles is already large (even in the one-way coupled case). At $d_p=18 \micron$, the deposition fraction is close to one for one-way, two-way and four-way coupled simulations (\autoref{fig:depositionWithDpMouth}).
As particle size decreases below $d_p < 10 \micron$, the influence of four-way coupling appears to decrease at $100 \microg$ (total deposition is 0.12, 0.094 and 0.074 for $d_p=10,8$ and $6 \micron$ respectively).
The influence of four-way coupling on upper-airway deposition (\autoref{fig:depositionDiff}b) is larger than total deposition (five different $d_p$ values with upper airway deposition difference larger than 0.1, compared to one for total deposition difference, \autoref{fig:depositionDiff}a). 
This occurs since the drug is most densely clustered in this region at the beginning of inhalation (\autoref{fig:log10AlphaP-viz}a,b) and begins to be dispersed more as the patient inhales further. 

\begin{figure}
    \centering
    \includegraphics[width=\linewidth]{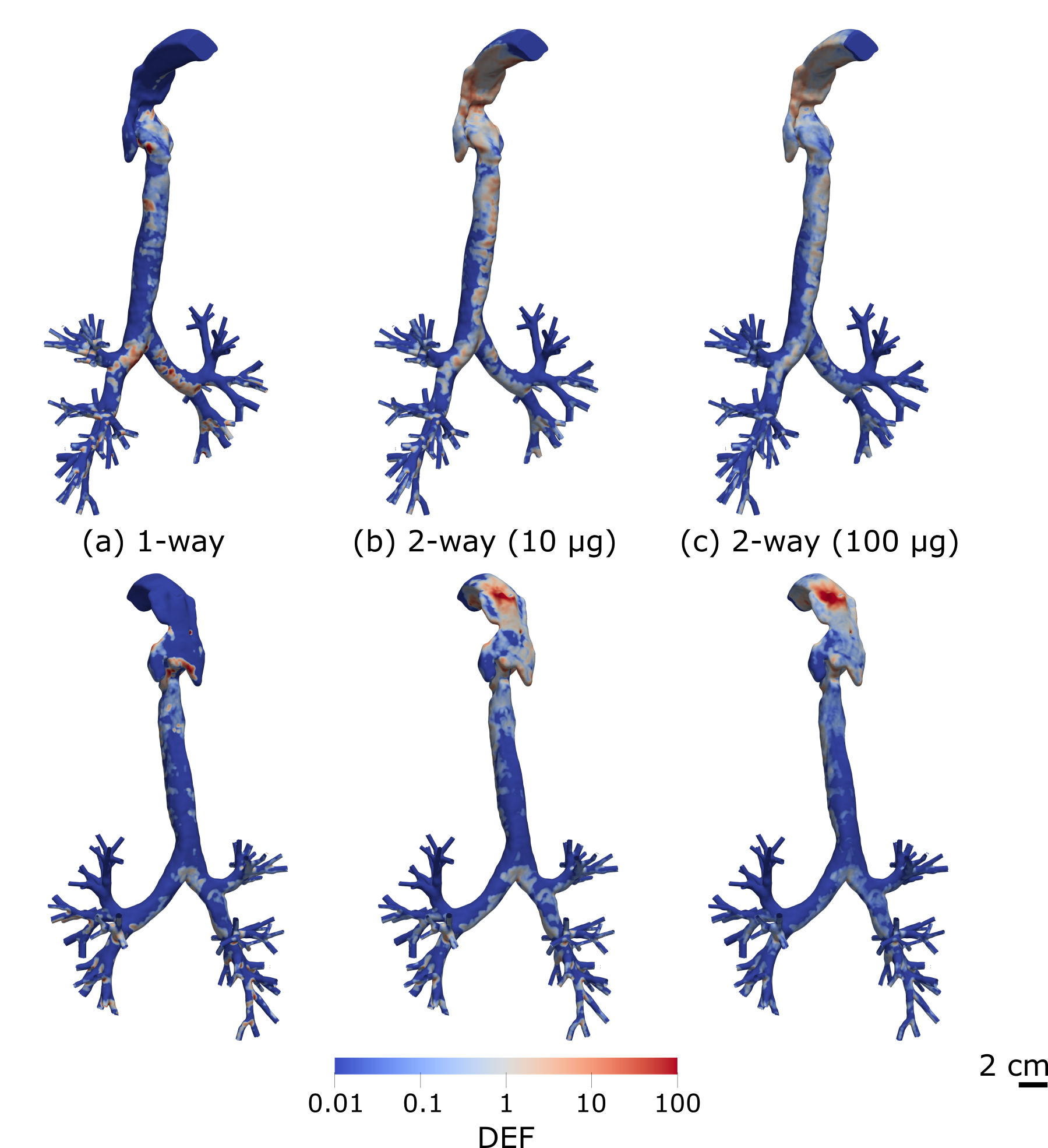}
    \caption{Deposition enhancement factor (DEF), Eq.~\eqref{eq:dosimetry}, of $10 \micron$ particles on the airway wall with various coupling and dosage parameters. Top row shows a front view, bottom row shows a read view. Panels show (a) 1-way coupling baseline, (b) 2-way coupling with $10 \microg$, (c) 2-way coupling with $100 \microg$.}
    \label{fig:depositionEF2way}
\end{figure}

\autoref{fig:depositionEF2way} shows the effect of two-way coupling on the DEF. We observed a clear difference between deposition fields in one-way and two-way coupling. In one-way coupling (\autoref{fig:depositionEF2way}a), deposition concentration is high at airway bifurcations as reported widely in literature such as \citet{kleinstreuer2010review}. There are also some patches of deposition in the trachea and mouth due to secondary flows here, as shown by \citet{williams2022effect} for $N_p=2\times 10^5$.
However, the deposition is uniform in the mouth and throat in the case of two-way coupling (\autoref{fig:depositionEF2way}b). 
From \autoref{fig:velocitynormalPDF-10micron}, we found that the effect of two-way coupling creates less correlated trajectories which leads to more uniform distribution and deposition of the particles (\autoref{fig:depositionEF2way}b,c). 
To verify this was not due to statistical bias in the limited number of particles typically used for one-way coupling in literature, we used the same $N_p$ in panels (a) and (b). 
Increasing dosage from $10 \microg$ to $100 \microg$ (\autoref{fig:depositionEF2way}c) shows that deposition is more uniform in the upper part of the mouth, deposition in the trachea and first bifurcation is lower. 

\section{Discussion}

In this study we performed large-scale simulations tracking a realistic number of aerosol drug particles in patient-specific airways to understand the influence of two-way and four-way coupling on deposition.
By examining variations in the distributions of physical quantities such as solid volume fraction, orientation between particle and fluid vectors, as well as collision frequency, we were able to explain the observed increase in deposition for two-way coupled simulations.
The two-way coupled deposition fractions were significantly larger than the one-way coupled results as modelled mass was increased. 
This is explained by a lower correlation between particle and fluid velocity vectors, which creates a more uniform particle concentration.
As the mass was increased towards $100 \microg$, regions of the flow fell into the four-way coupling regime ($\alpha_p > 10^{-3}$) which caused an increase in particle-phase pressure that drives particles towards low pressure (low $\alpha_p$) regions, and decreases total deposition by up to 12\% (relative to total dosage).
Therefore, we recommend deposition simulations include four-way coupling, which can be made computationally efficient using parcel-modelling.

To allow further study with realistic dosages with more modest computational resources, we investigated the effect of parcel modelling on deposition and other flow properties (\autoref{sec:results-parcel}). We observed very good agreement between the simulation with 10 particles per parcel ($N_{parcel}=10$) and our baseline ($N_{parcel}=1$), which was also used by \citet{kolanjiyil2021importance} for nasal sprays. 
The simulation data with 100 particles per parcel give a visually similar deposition enhancement factor (\autoref{fig:parcelDEF10um}), and deposition fraction error of below 10\%. However, the solid volume fraction distribution began to depart from the baseline at $t = 0.4 \jwunit{s}$ (\autoref{fig:coarse-solidVolPDF}c,d). 
Increasing the coarsening factor to 1000 particles per parcel yielded a maximum deposition error of 24\% relative to total dosage (\autoref{fig:parcel}). 
At $N_{parcel}=100$, the solid volume fraction and granular temperature distributions began to differ from the $N_{parcel}=1$ results (Figures \ref{fig:coarse-solidVolPDF} and \ref{fig:coarse-grantemp}).
Increasing $N_{parcel}$ to 1000 gave a much more significant change in the distributions of solid volume fraction and granular temperature, leading to a larger and intolerable deposition error (up to 23.9\%).
Alternatives to parcel modelling that could be explored are Eulerian methods such as solving a velocity density function \citep{patel2019three, li2022implementation}, where the computational cost is independent of $N_p$.

Four-way coupling was performed in our previous study \citep{williams2022effect}, as we used CFD-DEM to resolve particle-particle interactions. 
The high computational cost of DEM limited us to $N_p = 10^6$ (dosage $0.6 \microg$), which showed a deposition increase of only 2\% compared to $N_p = 20\times 10^5$ (close to one-way coupling).
Aside from the larger number of particles in this study, we also used a time-varying boundary condition from a healthy patient \citep{cola04analtidal}, compared to a constant flowrate in \citet{williams2022effect}.
The constant flowrate boundary condition caused particles to quickly be dispersed throughout the airways, where as particles remain clustered in the mouth-throat area due to the slow flowrate at the onset of inhalation in this study. 
These two factors explain the larger influence of two and four-way coupling on deposition observed in the present study.

This study focused on a single patient lung airway and breathing profile. Future efforts should investigate the effect of modelling realistic dosages in a broader range of patients, and consider different inhalation profiles.
We expect the influence of realistic dosages will be more significant in infants due to the smaller size of their airways which will cause the particles to be compacted into a tighter space (larger mass-loading).
We also expect slower and more gradual inhalations to show a more significant effect of realistic dosages than fast inhalations such as the `impaired` inhalation from our previous study \citep{williams2022effect, cola04analtidal}. 
Medical device manufacturers could mitigate the increased deposition we observed by designing devices with a slower release of particles (such as Soft Mist \textsuperscript{TM} \citep{hochrainer2005comparison}).

One limitation of this study is the use of the multiphase particle-in-cell (MPPIC) approach for modelling particle-particle collisions. Our previous study used the discrete element method (DEM) to track particle-particle interactions \citep{williams2022effect}.
However, this method imposed excessively small timesteps as the particle mass was decreased (which determines the collision duration). 
Therefore, we chose to use the MPPIC approach which replace the particle-particle collision with a particle pressure computed on the fluid grid \citep{andrews1996multiphase, snider2001}.
\citet{lu2017assessment} discuss MPPIC and alternative methods which can reduce computational cost (compared to DEM). 
A potential higher fidelity method than MPPIC which can use a larger timestep than DEM is a hard-sphere approach, where collisions occur over a single timestep (not 10-60 as is used in DEM).
However, the timestep size is still based on the collision duration, and therefore could become excessively small for $d_p = 1 - 5 \micron$.
A small number of hard-sphere simulations could be performed for larger particles with low dosages ($d_p \approx 10 \micron$ and $10 \microg$), to benchmark our MPPIC results.

\section{Conclusion}

We have performed large-scale simulations of drug inhalation in patient-specific airways with a realistic number of particles to study the effect of two/four-way coupling.
We used the simulation data to develop further understanding of the changes in flow physics in this case, finding that one-way coupling leads to an unphysical clustering of particles. In simulations with two-way coupling, the particles are decorrelated with the fluid and are more uniformly distributed. This leads to a higher total deposition fraction, as well as a more uniform spatial distribution of deposited particles.
Future deposition simulations should be fully-coupled, to avoid significant under-prediction of deposition caused by assuming one-way coupling.
This will allow for realistic predictions of multiphase flow physics required to inform clinical decision-making.

\section*{Acknowledgements}
Simulations reported in this study were performed on Oracle Cloud Infrastructure computing platform, funded by Open Clouds for Research Environments (OCRE) `Cloud Funding for Research'.
JW was funded by a 2019 PhD Scholarship from the Carnegie-Trust for the Universities of Scotland.

\begin{appendices}

\section{OpenFOAM modifications} \label{app:code}

By default, OpenFOAM counts all particles (deposited and floating) when calculating $\alpha_p$. 
In simulations with realistic dosages, we found that this caused numerical errors as some cells on the wall would contain a large number of deposited/inactive particles.
This would cause the source term in the fluid momentum equation to be very large and spurious here, and impose a very small timestep on the fluid solver.
In respiratory simulations deposited particles will begin to interact with the mucus layer \citep{feng2021tutorial}, which is not modelled in our study.
Therefore, they do not have any effect on the flow hydrodynamics.
To account for this, we modified the OpenFOAM library to account for only floating (called `active') particles when computing $\alpha_p$.
Studies on mucus layer mechanics have reported the mucus layer thickness as $0.5 - 1 \jwunit{mm}$ for COPD \citep{yi2021computational}, which justifies our assumption that deposited particles will be immersed in the mucus layer and not influence the flow.
The modified source code is available on GitHub and has been archived with Zenodo \citep{codeOCREpaper}.

\end{appendices}

\bibliography{refs}

\end{document}